\documentstyle[11pt]{article}
\begin{document}
\begin{flushright}SJSU/TP-96-14\\August 1996\end{flushright}
\vspace{1.7in}
\begin{center}\Large{\bf Consistent Histories May be Strange, \\
          But So is Standard Quantum Theory}\\
\vspace{1cm}
\normalsize\ J. Finkelstein\footnote[1]{
        Participating Guest, Lawrence Berkeley National Laboratory\\
        \hspace*{\parindent}\hspace*{1em}
        e-mail: FINKEL@theor3.lbl.gov}\\
        Department of Physics\\
        San Jos\'{e} State University\\San Jos\'{e}, CA 95192, U.S.A
\end{center}
\begin{abstract}
Interpretational questions that arise in the Consistent Histories
formulation of quantum mechanics are illustrated by the familiar example
of a beam passing through multiple slits.
\end{abstract}
\newpage
In the consistent histories formulation of quantum mechanics
\cite {Grif84,Omnes,GMH}, one classifies certain sets of histories as
being ``consistent sets'', and then identifies values of the decoherence
functional as being the ``probabilities'' of consistent histories.
The consistency conditions ensure that, within any single consistent
set, the values of probabilities so obtained do have the
expected properties.  However (as has been known from the 
beginning \cite{Grif84,DEsp}, and has recently been re-emphasized in
\cite{K1,K2}), one will not necessarily get sensible results if one
combines the probabilities of histories in different sets. Thus it is
not clear how to interpret these ``probabilities''.
   
Feynman has written \cite{Feyn} that ``the {\em only} mystery'' of
quantum theory is encountered in the example of a beam passing though
multiple slits.  Guided by this insight, in this paper we will discuss
such an example, to try to illustrate (alas, not to resolve) some of 
the interpretational questions which are raised in \cite{K1,K2}.

So we consider, as shown in the figure, a beam of particles incident
upon a wall in which there are three slits, which we denote by $S_1$, 
$S_2$, and $S_3$; each slit may be either open or closed, as we choose.
We will not be interested in the entire diffraction pattern, and so we
imagine that there is a single detector called $D$ at a fixed location
on the far side of the wall.  We will only consider those events in which   
$D$ registers the passage of a particle.

Define $A_1$ to be the amplitude at $D$, in the case in which $S_1$ is open
while $S_2$ and $S_3$ are both closed. The counting rate in this case
is proportional to $|A_{1}|^2$. Make analogous definitions for $A_2$ and
for $A_3$, and let $A_{123}\equiv A_{1}+A_{2}+A_3$.
The counting rate when all slits are open is then proportional
to $|A_{123}|^2$. In general 
$|A_{123}|^{2}\neq |A_{1}|^{2}+|A_{2}|^{2}+|A_{3}|^2$,    
and standard quantum theory does not allow us to say that the particle
did go through any particular slit.

If we wish, we can think of $S_1$ and $S_2$ as a single object, which
we will call $S_{12}$ (it might be helpful, although not necessary, to
imagine that there were  no separation between $S_1$ and $S_2$); below we
will  refer to this way of thinking as
``analysis $\alpha$''. Define $A_{12}\equiv A_{1}+A_2$; then
the counting rate when $S_{12}$ is open and $S_3$ is closed is
proportional to $|A_{12}|^2$, and $A_{123}=A_{12}+A_3$.

In general, there will be interference between $A_{12}$ and $A_3$,
and standard quantum theory does not allow us to say that the particle
did go either through $S_{12}$ or through $S_3$.  Consider, however, 
the special case in which $A_{12}=0$.  In this case, there is nothing
for $A_3$ to interfere with; $A_{12}=0$ means that, if $S_{12}$ is open
and $S_3$ is closed, the particle could not reach $D$. 
We would surely be tempted, in this special case, to conclude that 
any particle detected by $D$ must have gone through $S_3$.

Thus analysis $\alpha$ apparently enables us to say that, if $A_1+A_2=0$,
particles detected by $D$ must have gone through $S_3$.  Of course, instead
of analysis $\alpha$,
we could have thought of $S_2$ and $S_3$ as a single object
(call this analysis $\beta$), and then by an analogous argument have said that
if $A_2+A_3=0$, particles detected by $D$ must have gone through $S_1$.
Now consider the case $A_1=A_3=-A_2$; since $A_1+A_2=0$, we conclude from
analysis $\alpha$ that the particle went through $S_3$, and since
$A_2+A_3=0$, we conclude from analysis $\beta$ that the particle went
through $S_1$.  Could both conclusions be correct?

It seems that, when a particle could arrive at $D$ by any of three paths,
we can get into trouble by asserting that the particle must have followed
one of the paths, even in the case in which the amplitudes of the other two 
paths cancel. \footnote {Even if there were only two slits, there
could still be similar trouble. 
Suppose that $S_1$ were closed (so we need only consider $S_2$ and $S_3$),
and suppose further
that $A_2=0$; then we still could not conclude that a detected
particle went through $S_3$. This is because $S_2$ can be said to have
an upper part and a lower part, and (using an obvious notation) it could
be that $A_3=A_{2,upper}=-A_{2,lower}$; then the fact that 
$A_{3}+A_{2,lower}=0$ could equally-well lead us to
conclude that the particle went through $S_{2,upper}$.}
It should be understood that there is nothing that is mysterious going
on here {\em beyond} the mysteries of good old quantum interference.
However, those mysteries are so familiar to us that we often forget
to be amazed by them.
Feynman, in discussing the two-slit experiment \cite{Feyn}, stresses that,
because of quantum interference,
we can not say that particles arriving at $D$ went through either
of the two slits.  We now see that this remains true even when the 
amplitude for one of the slits vanishes.       

This example, of particles passing through multiple slits, can of course
be expressed in the formalism of consistent histories. Let the particle
initially be in a pure state; then, with the condition
$A_1=A_3=-A_2$, the set of four histories in which the particle goes through
either $S_{12}$ or $S_3$, and then is either detected by $D$ or is not,
is a consistent set. Consideration of this set corresponds to analysis
$\alpha$ discussed above. Within this set, the
history consisting of a particle going through $S_{12}$ and then being
detected has zero probability (which follows immediately from the condition
$A_{12}=0$, and is what ensures that the set is consistent), and so the 
conditional probability for a detected particle to have gone through $S_3$
has value one.  Likewise, the set of four histories in which the particle
goes through either $S_1$ or $S_{23}$, and then is either detected or not,
is also a consistent set (corresponding to analysis $\beta$).  
Within {\em this} set, the 
conditional probability for a detected particle to have gone through $S_1$
also has value one.  In fact, this example is one of the class of examples
considered in \cite{K1,K2}; the contradiction referred to in the title
of \cite{K1} is between the retrodiction that a detected particle went
through $S_3$ and the retrodiction that it went through $S_1$.  In \cite{K2}
this contradiction is expressed in a different way, which in our example
would be the observation that we might conclude (using analysis $\alpha$)
that a detected particle went through $S_3$, and also (using analysis $\beta$)
that it did {\em not} go through $S_{23}$.
Since $S_3$ is a part of $S_{23}$, one would have thought that going 
through $S_3$ implied going through $S_{23}$. 

If we want to apply the consistent history formalism to a closed system,
and {\it a fortiori} if we want to apply it to the universe, we do not
have the option of refusing to accept any conclusion that is not verified
by an external measuring device. We are thus faced with the question
of under what circumstances and in what sense we should understand the
``probabilities'' which are assigned to consistent histories. To answer
this question, several authors \cite{Omnes,Grif96,Ish} have 
proposed versions of logic in which statements would only have validity
in the context of a given analysis.  In our example, this would amount
to accepting, for a detected particle,
{\em both} statements ``In analysis $\alpha$, it
went through $S_3$'' and ``In analysis $\beta$, it went through
$S_1$''. Then, if someone were to say to us ``Now that you have made both
analyses, please tell me through which slit the particle {\em really did} 
go,'' we would refer him to \cite{Feyn}.

\vspace{1cm}
Acknowledgement:
I would like to acknowledge the hospitality of the
Lawrence Berkeley National Laboratory.

\vfill
\setlength{\unitlength}{0.1in}
\begin{picture}(45,40)
\thicklines
\put(0,24){\line(1,0){5.5}}
\put(0,24.5){\line(1,0){5.5}}
\put(24,29){\line(0,1){8}}
\put(24.1,29){\line(0,1){8}}
\put(24,24){\line(0,1){4}}
\put(24.1,24){\line(0,1){4}}
\put(24,19){\line(0,1){4}}
\put(24.1,19){\line(0,1){4}}
\put(24,10){\line(0,1){8}}
\put(24.1,10){\line(0,1){8}}
\thinlines
\put(25,29){$S_1$}
\put(25,24){$S_2$}
\put(25,19){$S_3$}
\put(42,28){\framebox(3,3){}}
\put(44,32){$D$}
\put(2,27){\em Beam}
\put(12.5,5){A beam passing through three slits}
\Huge
\put(4.2,23.5){$>$}
\end{picture}
\end{document}